# Domain-wall driven suppression of thermal conductivity in a ferroelectric polycrystal

*Rachid Belrhiti-Nejjar[1,\*], Manuel Zahn[2,3\*], Patrice Limelette[1], Max Haas[2,4], Lucile Féger[1], Isabelle Monot-Laffez[1], Nicolas Horny[5], Dennis Meier[2], Fabien Giovannelli[1,\*\*], Jan Schultheiß[2,6,\*\*\*], Guillaume F. Nataf[1,\*\*\*\*]*

[1] GREMAN UMR 7347, Université de Tours, CNRS, INSA-CVL, 16 rue Pierre et Marie Curie, 37071 Tours, France
[2] Department of Materials Science and Engineering, NTNU Norwegian University of Science and Technology, Høgskoleringen 1, Trondheim 7034, Norway
[3] Experimental Physics V, Center for Electronic Correlations and Magnetism, University of Augsburg, 86159 Augsburg, Germany
[4] German Aerospace Center (DLR), Institute of Materials Research, 51147 Cologne, Germany
[5] Institut de Thermique, Mécanique et Matériaux, Université de Reims Champagne-Ardenne, 51687 Reims, France
[6] Department of Mechanical Engineering, University of Canterbury, 8140 Christchurch, New Zealand
\* equal contributions; \*\* fabien.giovannelli@univ-tours.fr; \*\*\* jan.schultheiss@ntnu.no; \*\*\* guillaume.nataf@univ-tours.fr

A common strategy for reducing thermal conductivity of polycrystalline systems is to increase the number of grain boundaries. Indeed, grain boundaries enhance the probability of phonon scattering events, which has been applied to control the thermal transport in a wide range of materials, including hard metals, diamond, oxides and 2D systems such as graphene. Here, we report the opposite behavior in improper ferroelectric $ErMnO_3$ polycrystals, where the thermal conductivity decreases with increasing grain size. We attribute this unusual relationship between heat transport and microstructure to phonon scattering at ferroelectric domain walls. The domain walls are more densely packed in larger grains, leading to an inversion of the classical grain-boundary-dominated transport behavior. Our findings open additional avenues for microstructural engineering of materials for thermoelectric and thermal management applications, enabling simultaneous control over mechanical, electronic, and thermal properties.

## 1. Introduction

Materials with ultra-low thermal conductivity[1] are essential for various applications, ranging from efficient thermoelectric devices[2, 3] and thermal barrier coatings[4] to insulation in cryogenic systems[5]. In electrically insulating solids, strategies to reduce thermal conductivity involve hindering of phonon transport, i.e., lattice vibrations, which are the primary heat carrier.[6] An established approach for decreasing the mean free path for phonons and, hence, reducing the heat flow, is to increase the number of interfaces between different phases (hetero-interface) or grains (grain boundary)[2] at which the phonons scatter. A typical example is polycrystalline thermoelectric skutterudite, where the grain boundaries reduce the thermal conductivity by about an order of magnitude compared to the single crystalline counterpart[7]. The same trend is observed in oxides (e.g., tin oxide[8, 9] and aluminum-doped zinc oxide[10]), where the thermal conductivity can be set to values between ~1 W m$^{-1}$ K$^{-1}$ and ~40 W m$^{-1}$ K$^{-1}$, controlled by the grain boundary volume fraction. The effect extends to high thermal conductive polycrystalline materials, such as hard metals[11] or diamond[12]. Even in two-dimensional materials, e.g., graphene, the thermal conductivity decreases with decreasing grain size,[13] indicating that the grain-size-dependent scaling of thermal conductivity is largely material-independent.

Another type of interface that strongly interacts with phonons are ferroelectric domain walls[14-22]. Ferroelectric domain walls are natural interfaces that arise in system with long-range order electric dipoles, separating regions with different polarization orientation. Domain wall–phonon interactions become particularly strong when the distance between the walls approaches the mean free path of the phonons[14-16], which is around 10 nm to 100 nm at room temperature in ferroelectric or ferroelastic oxides[14, 17]. For this reason, related work mainly focuses on nanoscale domains in films with a thickness in the sub-micrometer range[15, 16]. One intriguing example of a bulk material where a correlation between the domain wall density and thermal transport behavior was reported is improper ferroelectric $ErMnO_3$[23]. The thermal conductivity of single crystals was observed to decrease along with the size of the domains, indicating the importance of the domain walls for heat transport. This observation motivates our work on $ErMnO_3$ polycrystals, enabling a systematic study of the concerted impact of coexisting ferroelectric domain walls and grain boundaries.

Here, we synthesize $ErMnO_3$ polycrystals with varying grain and ferroelectric domain sizes to quantify the different contributions from grain boundaries and ferroelectric domain walls on the thermal conductivity. We observe a pronounced decrease in thermal conductivity as the domain size decreases, indicating that domain walls efficiently suppress thermal conductivity. Importantly, we find that domain walls overrule grain boundary effects in $ErMnO_3$, which leads to an inversion of the well-established grain-size-dependent behaviour of thermal conductivity in solids. Our findings show how domain wall engineering can be leveraged to control the thermal conductivity in a polycrystalline ferroelectric material, giving additional opportunities for achieving systems with ultra-low thermal conductivity.

## 2. Tailoring thermal conductivity through domain engineering

We begin by determining the basic crystallographic and microstructural properties of our $ErMnO_3$ polycrystals, which are essential for their thermal transport behavior. X-ray diffraction (XRD) confirms that all polycrystals have the same space group symmetry as ErMnO$_3$ single crystals ($P6_3cm$[24], Fig. S1a), independent of the applied cooling rate. The relative geometrical densities of the samples are between 89% and 92% (Table 1) as confirmed by scanning electron microscopy (SEM) images (Fig. S4). SEM images reveal microcrack formation caused by thermal expansion anisotropy. We note that the crack formation does not correlate with the cooling rate as cracks form below $T_C$,[25] where all samples were cooled at the same rate. Furthermore, due to the identical cooling conditions below $T_C$, we can exclude pronounced sample-to-sample variations in oxygen stoichiometry, which is predominantly controlled by dwelling times at around 350°C[26]. Grain sizes, obtained by averaging over ~50−75 grains for each cooling rate (Tab. 1, Fig. S2 and Supplementary Information S4), are in the range of 9.9±0.1 μm (0.01 °C/min) to 8.3±0.1 μm (10 °C/min) (Tab. 1). Crucially, while the grain size decreases by ≈16% with increasing cooling rate, samples cooled at 0.1°C/min, 1°C/min, and 10°C/min show nearly identical grain sizes, differing by less than 5%.

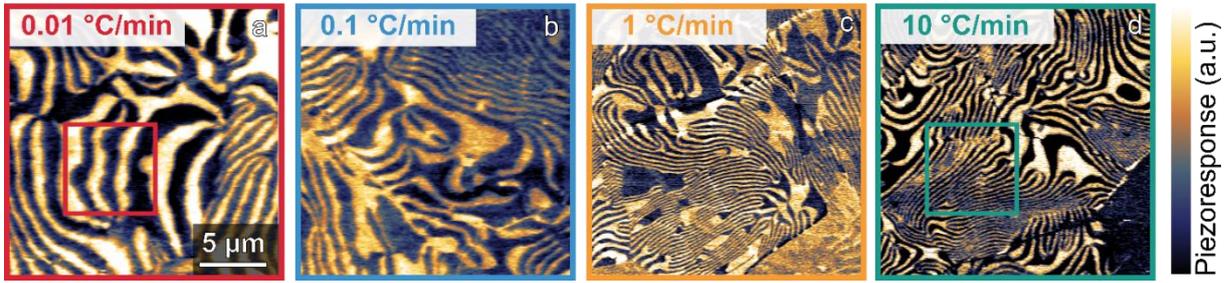

**Figure 1. Influence of cooling rate variation on ferroelectric domain formation.** PFM images of samples cooled under a) 0.01 °C/min, b) 0.1 °C/min, c) 1 °C/min and d) 10 °C/min in the temperature window between 1176 °C and 1136 °C in the vicinity of $T_c$. Data was recorded with a peak-to-peak excitation voltage of 10 V at a frequency of 40.13 kHz. Light and dark regions correspond to ferroelectric domains with opposite directions of polarization $\pm P$. Topography data obtained on the same area is displayed in Fig. S2. A magnified view of the domain structures highlighted by the squares in panels a) and d) is shown in Fig. 2b.

Figure 1 shows representative piezoresponse force microscopy (PFM) images of our polycrystalline ErMnO$_3$ samples. ErMnO$_3$ is a geometrically driven improper ferroelectric, where ferroelectric polarization, $P$, arises from a structural trimerization mode[27-29]. As a uniaxial ferroelectric with $P$ parallel to the hexagonal $c$-axis ($P \approx 6\,\mu C/cm^2$), its domain structure consists of 180° ferroelectric domains that converge at topologically protected six-fold structural vortex lines as described elsewhere.[30] We observe a pronounced PFM contrast that allows for distinguishing the respective +*P* and -*P* domains[31], showing the characteristic domain structure of polycrystalline hexagonal ErMnO$_3$[24].

**Table 1.** Relative geometrical density, grain size, domain size, and thermal conductivity at room temperature. Relative geometrical densities are calculated using 7.286 g cm$^{-3}$ as theoretical density[33].

| Heat treatment conditions | Relative density [%] | Grain size [μm] | Domain size [nm] | Conductivity [W m$^{-1}$ K$^{-1}$] |
|---|---|---|---|---|
| *Cooling rate variation: Domain size changes, grain size fixed* | | | | |
| 0.01 °C/min | 92 | 9.9±0.1 | 715 | 3.0 |
| 0.1 °C/min | 90 | 8.8±0.1 | 460 | 2.7 |
| 1 °C/min | 89 | 8.5±0.2 | 419 | 2.5 |
| 10 °C/min | 91 | 8.3±0.1 | 386 | 2.2 |
| *Grain size variation: Grain size changes, domain size changes* | | | | |
| 1350 °C, 10 min | 85 | 1.2±0.1 | 351 | 3.6 |
| 1350 °C, 4 hrs | 89 | 3.0±0.1 | 333 | 3.5 |
| 1450 °C, 12 hrs | 92 | 8.4±0.2 | 274 | 3.2 |

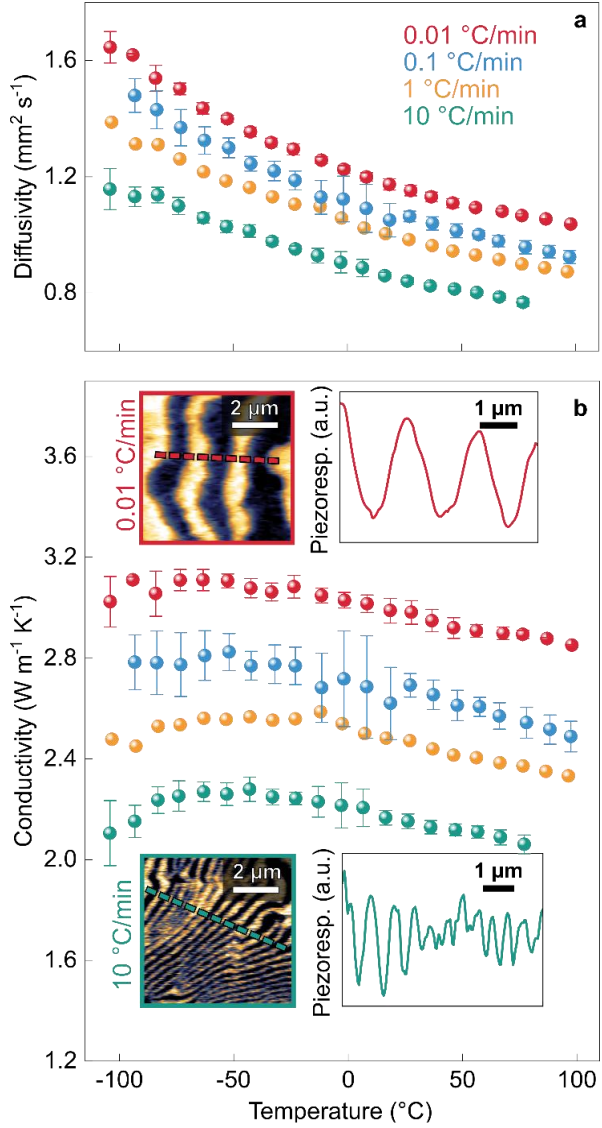

**Figure 2: Influence of cooling rate variation on thermal conductivity.** a) Thermal diffusivity and b) thermal conductivity as a function of temperature for samples cooled at different rates (recorded on heating). At each temperature, three measurements were taken, with standard deviations represented as error bars. For the sample cooled at 1 °C/min, only a single measurement was performed. Carbon coating was not required, as the samples inherently absorbed light from the 650 nm laser, eliminating potential errors from variations in coating thickness. Thermal diffusivity was extracted by fitting the laser flash data to the Cape-Lehman model[32]. Across the entire temperature range, the thermal conductivity decreases by approximately 25%. Representative PFM images illustrate the ferroelectric domain structure of the samples cooled at the fastest (10 °C/min) and slowest (0.01 °C/min) rates, corresponding to the areas highlighted in Fig. 1a and d, respectively. Line plots extracted along the dashed lines visualize the dependence of domain periodicity on the cooling rate.

Notably, we observe a significant cooling-rate dependence with the domain size decreasing as the cooling rate increases.

Thermal diffusivity was measured on the polycrystals via the laser flash method (LFA 457, Netzsch, Germany) in a temperature range from -105°C to +96°C (Fig. 2a). The four samples exhibit a decrease of thermal diffusivity with increasing temperature with sample-dependent absolute values. Subsequently, the diffusivities and geometrical densities are used to compute temperature-dependent thermal conductivities (Fig. 2b), using heat-capacity data from ref. [23]. At all temperatures, the thermal conductivity depends on the cooling rate through the phase transition temperature during processing. Thereby, higher cooling rates lead to greater suppression of the thermal conductivity, with a ~25% higher thermal conductivity measured for the sample cooled under the smallest (0.01 °C/min) compared to the largest (10 °C/min) rate. Consistent with previous findings on single crystals of ErMnO$_3$,[23] we attribute the suppression of thermal conductivity with cooling rate in our polycrystals to the formation of ferroelectric domains and domain walls. The correlation between ferroelectric domain structure and thermal conductivity is displayed by representative PFM images as insets in Fig. 2b for the two end cases (0.01 °C/min and 10 °C/min), with the distance-dependent piezoelectric response extracted along the dashed lines. The extracted line-profiles highlight the decrease in domain size with increasing cooling rate. A detailed analysis (Supplementary Information S4, Tab. 1), averaging 50–75 grains for each sample, confirms this trend, with the slowest-cooled sample (0.01 °C/min, Fig. 1a) exhibiting the largest mean domain size (~715 nm) and the fastest-cooled sample (10 °C/min, Fig. 1d) the smallest (~386 nm). Similar to the findings on single crystals of ErMnO$_3$,[23] the control of thermal conductivity in our series remains effective at room temperature, with domain walls reducing the thermal conductivity from 3.0 (0.01 °C/min) to 2.2 W m$^{-1}$ K$^{-1}$ (10 °C/min). The results extend previous studies towards polycrystalline samples where domain walls coexist with additional grain boundaries, which we will explore in the next step.

### 3. Domain-wall driven suppression of thermal conductivity

To disentangle the contributions from domain walls and grain boundaries, we synthesize three polycrystalline samples with different grain sizes from the same powder. This is done by varying temperatures and dwell times from 1350 °C, 10 min to 1350 °C, 4 hrs and 1450 °C, 12 hrs.[24] For all samples, heating and cooling are performed at 5 °C/min. XRD confirms the *P*6$_3$*cm* space group symmetry (Fig. S1b). The relative geometrical densities of samples are 85% (1350 °C, 10 min), 89% (1350 °C, 4 hrs) and 92% (1450 °C, 12 hrs), as summarized in Tab. 1. The size of the grains is extracted from PFM data (Fig. S3). Averaging over ~15−90 grains for each cooling rate, we find a pronounced variation in grain size with sizes of 1.2±0.1 µm (1350 °C, 10 min), 3.0±0.1 µm (1350 °C, 4 hrs) to 8.4±0.2 µm (1450 °C, 12 hrs). Figure 3 shows representative PFM images, with all samples featuring a mixture of vortex and stripe-like domains. In the sample heat treated at 1350 °C for 10 min, the formation of the vortex- and stripe-like domain structure is largely suppressed (Fig. 3a), i.e., the grains approach a single domain state consistent with ref. [24]. Importantly for this work, the heat treatment conditions have a significant effect on the ferroelectric domain structure, with the smallest domains observed in the sample treated at 1450 °C for 12 hrs. This finding is in agreement with previous observations on hexagonal manganite polycrystals,[24, 34] where the reduction in domain size for larger grain sizes was attributed to the interaction of topological vortex lines with strain fields, facilitating the transformation into the observed stripe-like configurations.

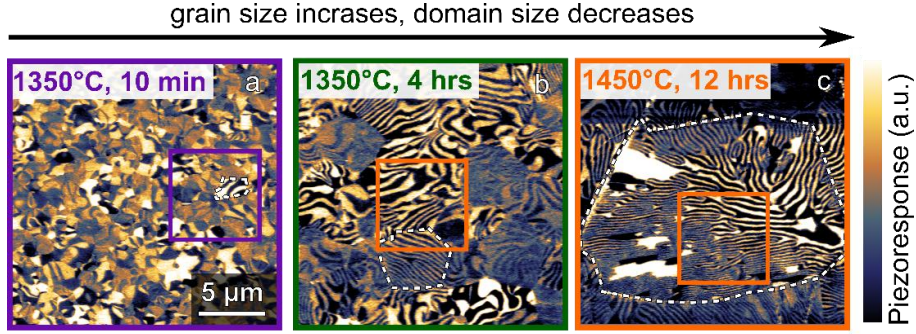

**Figure 3: Influence of varying temperature and dwell time on ferroelectric domain formation.** PFM images of samples heat treated under different temperature and dwell times, a) 1350 °C, 10 min, b) 1350 °C, 4 hrs, and c) 1450 °C, 12 hrs. Dashed white lines indicate the position of grain boundaries. Light and dark regions correspond to ferroelectric domains with opposite directions of the polarization ±*P*. Topography data obtained on the same area is displayed in Fig. S3. A magnified view of the domain structures highlighted by the squares in panels a) and d) is shown in Fig. 4b.

Thermal diffusivity (Fig. 4a) and thermal conductivity (Fig. 4b) are measured for the three polycrystalline samples using the same procedure as described above. The samples exhibit distinct thermal conductivities, with the highest value (3.6 Wm$^{-1}$K$^{-1}$ at room temperature) measured for the sample treated at 1350 °C for 10 min. In contrast, the sample treated at 1450 °C for 12 hrs shows a ~12% lower thermal conductivity of 3.2 Wm$^{-1}$K$^{-1}$. The variation in absolute thermal conductivity values between the two sample series (Fig. 2 and Fig. 4) is attributed to differences in heat treatment conditions (see Supplementary Information S1 for more details).

The relationship between thermal conductivity and grain size is displayed in Fig. 5. Unlike in conventional polycrystalline materials, we find that the thermal conductivity in ErMnO$_3$ decreases with increasing grain size. This inverse trend is in contrast to literature data on, e.g., polycrystalline SnO$_2$,[9] Al-ZnO,[10] and WC-Co[11], where thermal conductivity increases with grain size. This unusual behavior suggests that the grain-size dependence of thermal conductivity is governed by a mechanism other than the conventional phonon scattering at grain boundaries. Based on the results reported for single crystals[18], and our data gained on polycrystalline samples (Fig. 1 and 2), we conclude that this behavior originates from the varying domain wall density (see insets and line profiles in Fig. 4b). The line profiles reflect that the sample treated at the highest temperature (1450 °C for 12 hrs) has a smaller domain size compared to the sample treated at the lowest temperature (1350 °C for 10 min). A more detailed analysis (Tab. 1) confirms this trend, with mean domain sizes of 351 nm (1350 °C, 10 min), 333 nm (1350 °C, 4 hrs), and 274 nm (1450 °C, 12 hrs). For reference, our results are plotted together with grain-size-dependence thermal conductivity data we gained on the model ferroelectric BaTiO$_3$ (grain size range of 0.1 to 50 µm). BaTiO$_3$ shows the classical relation expected for polycrystals[9-11, 35], and the comparison between as-grown BaTiO$_3$ and ErMnO$_3$ emphasizes the unusual behavior of our ErMnO$_3$ polycrystals.

To quantify for the contributions, we apply a one-dimensional heat transfer model, where the thermal resistivity of the samples is described as the sum of resistances in series, corresponding to the grain boundaries and domain walls, and the intrinsic thermal resistance of the material $R_i = e/\kappa_i$, with $e$ being the thickness of the sample and $\kappa_i$ the intrinsic thermal conductivity. The effective thermal conductivity is thus expressed as:

$$\kappa = \frac{1}{\frac{R_{GB}}{l_{GB}} + \left(\frac{1}{l_{DW}} - \frac{1}{l_{GB}}\right)R_{DW} + \frac{1}{\kappa_i}},$$

where $l_{GB}$ and $l_{DW}$ are the grain and domain sizes, $1/l_{GB}$ and $1(1/l_{DW} - 1/l_{GB})$ multiplied by $e$ is the number of grains and domains, respectively. $R_{GB}$ and $R_{DW}$ are the respective Kapitza resistances[36], measuring the resistance to thermal flow at the interface between two grains and two domains, respectively.

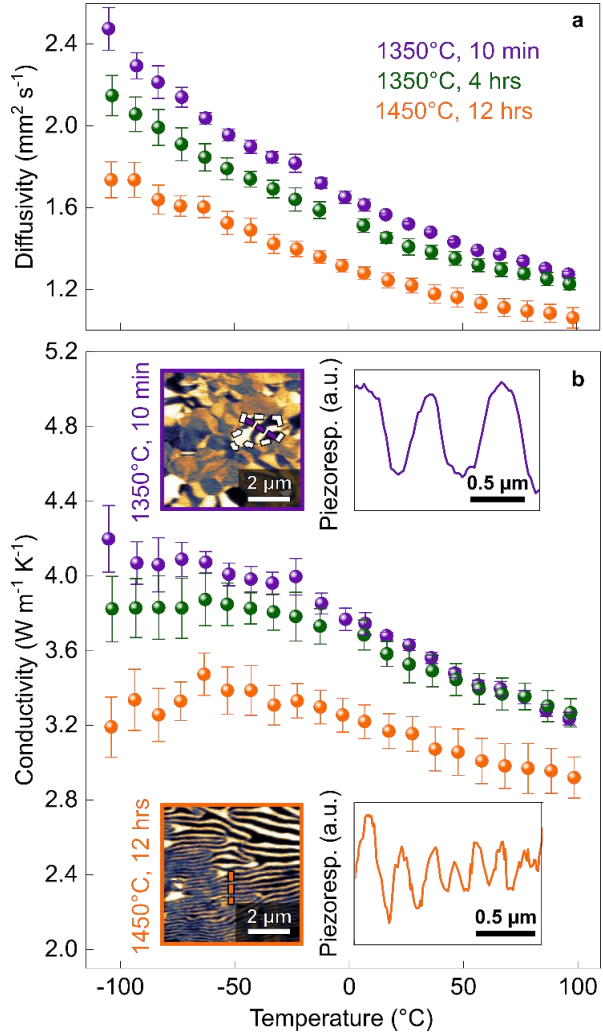

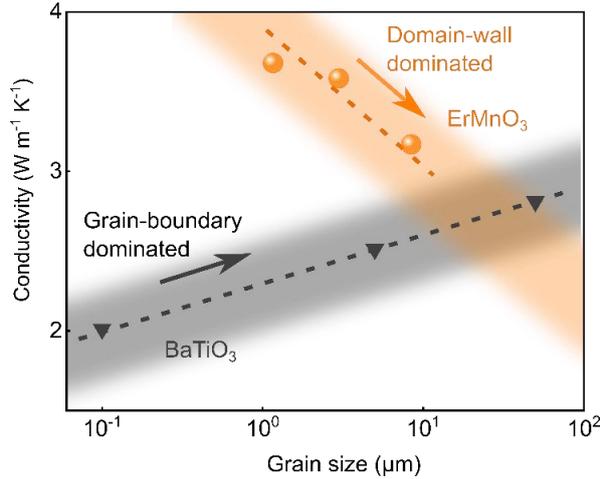

**Figure 4: Influence of temperature and dwell time on thermal conductivity.** a) Thermal diffusivity and b) thermal conductivity as a function of temperature for samples heat-treated under different temperatures and dwell times (recorded on heating). Three measurements were taken at each temperature, with standard deviations shown as error bars. Representative PFM images show the ferroelectric domain structure of samples heat-treated under two extreme conditions: 1350 °C for 10 min and 1450 °C for 12 hrs, corresponding to the highlighted areas in Fig. 3a and 3c, respectively. Line plots extracted along the dashed lines illustrate the relationship between domain periodicity and heat-treatment conditions.

**Figure 5. Grain-size-dependence of thermal conductivity.** Grain size-dependence for thermal conductivity at room temperature is displayed for $BaTiO_3$ and $ErMnO_3$. For $BaTiO_3$, an increase of thermal conductivity with grain size is found, indicating a common grain-boundary-dominated thermal conductivity. In comparison, the grain-size-dependence of the thermal conductivity for polycrystalline $ErMnO_3$ decreases with increasing grain size, indicating the important role of the domain walls for thermal conductivity control. The dashed lines indicate a guide to the eye.

Assuming that the samples have a similar $\kappa_i$, the resistances $R_{GB}$ and $R_{DW}$ can be estimated at each temperature by minimizing a cost function built as the sum of the square of the difference between each $\kappa_i$ calculated using eq. 1. Utilizing experimentally obtained values $\kappa$ from thermal flash measurements (Figs. 2 and 4), and $l_{GB}$ and $l_{DW}$ from PFM measurements (Figs. 1 and 3), the temperature-dependent thermal conductivity with a sample-independent $\kappa_i$ can be consistently described (Fig. S6). This thermal model allows for estimating thermal resistances between grains and domains, with mean values $R_{GB} = 1.6 \cdot 10^{-9}$ m²KW$^{-1}$ and $R_{DW} = 2.8 \cdot 10^{-8}$ m²KW$^{-1}$. These values align well with interfacial thermal resistances found in other oxides, such as $5 \cdot 10^{-9}$ m²KW$^{-1}$ for grain boundaries in $SrTiO_3$ polycrystals.[37] Thermal resistances for domain walls fall within a similar range of $5 \cdot 10^{-9}$ m²KW$^{-1}$ for $PbTiO_3$[16] and $2 \cdot 10^{-8}$ m²KW$^{-1}$ for $BiFeO_3$[22]. Importantly, the model corroborates that both ferroelectric domain walls and grain boundaries have a substantial impact on the thermal conductivity in our $ErMnO_3$ polycrystals. The difference in resistivities ($R_{DW} > R_{GB}$) is consistent with our experimental observations and indicates that the domain walls suppress thermal conductivity more efficiently than the grain boundaries. Possible scattering centers for the phonons include polarization discontinuities at the ferroelectric domain walls as well as point defects, which segregate towards domain walls in hexagonal manganites[38]. Future work may provide additional insight into the phonon scattering mechanism, e.g., by targeted manipulation of oxygen defects through annealing experiments[26].

## 4. Conclusion

In contrast to classical polycrystalline materials – where grain boundaries govern the grain-size dependence of thermal conductivity – our study demonstrates an inverted behavior in $ErMnO_3$ polycrystals. Through a comprehensive analysis of samples with similar grain sizes but different domain sizes, we show that ferroelectric domain walls effectively suppress thermal conductivity by impeding phonon transport. Most interestingly, this effect is evident as an observed decrease in thermal conductivity with increasing grain size, where the influence of domain walls overrides the decreasing grain boundary scattering effects. To quantify this unusual behavior, we calculated the Kapitza resistance, which was found to be higher for the domain walls than for the grain boundaries, corroborating the dominant effect of domain walls on the thermal conductivity of $ErMnO_3$ polycrystals.

The inversion of thermal conductivity with grain size is not limited to $ErMnO_3$ and can be extended for other polycrystalline materials with similar domain-size/grain-size scaling behavior. Potential candidates include isostructural hexagonal indates[39], tungsten bronzes[40], and 2H compounds[41]. The impact of ferroelectric domain walls on the grain-size dependence of thermal conductivity yields an innovative approach for tuning thermal properties. Our findings introduce an additional lever for adjusting thermal conductivity beyond traditional grain size modulation, opening an avenue for designing microstructures with tailored thermal behaviors. Properties such as fracture toughness, which typically improve with grain size,[42] could be optimized for ultra-low thermal conductivity applications. Gradient grain size distributions[43] further allow for localized control, providing a microstructure-related parameter that may be leveraged to generate inhomogeneous heat flow in electronics and thermoelectrics.

**Supplementary Information**
Supplementary Information is available from the Authors.

**Acknowledgements**
This work was co-funded by the European Union (ERC, DYNAMHEAT, N°101077402). Views and opinions expressed are, however, those of the authors only and do not necessarily reflect those of the European Union or the European Research Council. Neither the European Union nor the granting authority can be held responsible for them. J.S. acknowledges the support of the Alexander von Humboldt Foundation through a Feodor-Lynen research fellowship and the German Academic Exchange Service (DAAD) for a Post-Doctoral Fellowship (short-term program). M.Z. acknowledges funding from the Studienstiftung des Deutschen Volkes via a doctoral grant and the State of Bavaria via a Marianne-Plehn scholarship. D.M. thanks the NTNU for support through the Onsager Fellowship Program, the outstanding Academic Fellow Program. D.M and J.S. acknowledge funding from the European Research Council (ERC) under the European Union's Horizon 2020 Research and Innovation Program (Grant Agreement No. 863691).

**Author Contributions**
GFN, FG, JS, and DM conceived the study. RB-N performed thermal conductivity measurements on $ErMnO_3$ polycrystals and analyzed the data, under the supervision of GFN and FG. MH and JS synthesized the $ErMnO_3$ polycrystals. LF synthesized the $BaTiO_3$ polycrystals, under the supervision of IM-L, and measured their thermal conductivity under the supervision of IM-L and GFN. PL and NH developed the Kapitza model. MZ developed the procedure for domain and grain size quantification, supervised by JS. GFN and JS wrote the manuscript together with DM. All authors contributed to the discussion of the results and the final version of the manuscript.

**Conflict of Interests**
The authors declare no conflict of interests.

**Data Availability Statement**
The data that supports the findings of this study are available from the corresponding author upon reasonable request.

**Keywords**
Polycrystals, thermal conductivity, grain size, topological defects, domain walls

# Supporting Information

# Domain-wall driven suppression of thermal conductivity in a ferroelectric polycrystal

Rachid Belrhiti-Nejjar[1,*], Manuel Zahn[2,3*], Patrice Limelette[1], Max Haas[2,4], Lucile Féger[1], Isabelle Monot-Laffez[1], Nicolas Horny[5], Dennis Meier[2], Fabien Giovannelli[1,**], Jan Schultheiß[2,6,***], Guillaume F. Nataf[1,****]

[1] GREMAN UMR 7347, Université de Tours, CNRS, INSA-CVL, 16 rue Pierre et Marie Curie, 37071 Tours, France

[2] Department of Materials Science and Engineering, NTNU Norwegian University of Science and Technology, Høgskoleringen 1, Trondheim 7034, Norway

[3] Experimental Physics V, Center for Electronic Correlations and Magnetism, University of Augsburg, 86159 Augsburg, Germany

[4] German Aerospace Center (DLR), Institute of Materials Research, 51147 Cologne, Germany

[5] Institut de Thermique, Mécanique et Matériaux, Université de Reims Champagne-Ardenne, 51687 Reims, France

[6] Department of Mechanical Engineering, University of Canterbury, 8140 Christchurch, New Zealand

* equal contributions
** fabien.giovannelli@univ-tours.fr
*** jan.schultheiss@ntnu.no
*** guillaume.nataf@univ-tours.fr

## S1. Synthesis and characterization of polycrystalline ErMnO₃ and BaTiO₃

ErMnO$_3$ powder was obtained by a solid-state reaction between Er$_2$O$_3$ (99.9% purity; Alfa Aesar, MA, USA) and Mn$_2$O$_3$ (99.0% purity; Sigma-Aldrich, Germany). Powders were dried at 900°C and 700°C for 12 hours, mixed in a stoichiometric ratio and ball milled (BML 5, witeg Labortechnik GmbH, Germany) for 12 hrs at 205 rpm using yttria stabilized zirconia milling balls with a diameter of 5 mm and ethanol as dispersion medium. After drying, the resulting powder was annealed at 1000°C, mortared, annealed at 1050°C, mortared again and finally annealed at 1100°C, every time for 12 hrs. To study the impact of domain walls on thermal conductivity in polycrystals of ErMnO$_3$, powder was synthesized via a solid-state reaction followed by densification into bulk polycrystals as described in ref. [1].

To obtain dense materials similar in grain size, heat treatment was carried out at 1475°C for 24 hrs, with a heating and cooling rate of 5°C/min. Ferroelectric domain size control was achieved via variations in the cooling rates between 0.01 and 10°C/min (0.01, 0.1, 1, and 10°C/min) in the temperature range of 1176°C and 1136°C, near the ferroelectric transition, $T_c$ = 1156°C,[2] following established procedures described in refs. [2, 3] for single crystals. For realizing the different cooling rates (Fig. 1 and 2), sintering was carried out in a tube furnace (ETF 17 horizontal tube furnace, Entech, Ängelholm, Sweden). Before laser flash measurements, these samples were lapped with a 9-µm-grained $Al_2O_3$ water suspension (Logitech Ltd, Glasgow, Scotland) followed by polishing with silica slurry (SF1 Polishing Fluid, Logitech AS, Glasgow, Scotland).

To obtain the series with different grain sizes (Fig. 3 and 4), sintering was carried out in a box furnace (Entech Energiteknik AB, Ängelholm, Sweden) at temperatures ranging between 1350°C and 1450°C. Before laser flash measurements, these samples were mechanically polished. Thermal diffusivity was measured with a laser flash setup (LFA 457, Netzsch, Germany). Relative densities were obtained geometrically and by the Archimedes method (MS204TS/00 analytical balance, Mettler Toledo, Switzerland) in distilled water.

The $BaTiO_3$ polycrystals were synthesized from commercial $BaTiO_3$ nano- (purity 99.99%, Chempur, Poland) and micro-powders (purity 99.5%, Sigma-Aldrich, Germany), mixed with 5 % polyvinyl alcohol (2 wt.% in water, VWR International, USA), and then densified by conventional sintering at 1500 °C for different dwell times to obtain mean grain sizes of 5 µm (1400°C, 1 hrs) and 50 µm (1400°C, 6 hrs). A polycrystal with an average grain size of 0.1 µm was obtained by spark plasma sintering using a SPS632Lx (Dr.Sinter, Fuji Electronics, Japan). Sintering was performed with a graphite mold at 1050 °C for 3 min with a heating rate of 100 °C min$^{-1}$ under a uniaxial pressure of 100 MPa. Subsequently, the pellet was post-annealed for 1 hour under oxygen flow at 1000 °C in a tubular furnace to compensate for oxygen losses due to reducing conditions in the spark plasma sintering process. Thermal diffusivity was measured with the same laser flash setup as for the $ErMnO_3$ polycrystal (LFA 457, Netzsch, Germany). Silver glue paint and a thin graphite coating were used to maximize the infrared absorption and emission. The specific heat capacity was measured near room temperature on crushed polycrystals by differential scanning calorimetry (STA 449 F3 Jupiter, Netzsch, Germany) in platinum crucibles and nitrogen atmosphere. The relative density of the three polycrystals of $BaTiO_3$, obtained by the Archimedes method, were: 94% (0.1 µm grains), 97% (5 µm grains) and 94% (50 µm grains) considering a theoretical density of 6.02 g cm$^{-3}$.

## S2. Crystallographic analysis of ErMnO₃ polycrystals

X-Ray Diffraction (XRD, D8 ADVANCE, Bruker, MA, USA) of the investigated ErMnO$_3$ polycrystals is shown in Fig. S1. The crystallographic structures and phases are observed to be independent of the heat-treatment conditions and can be described with a hexagonal structure with space group symmetry $P6_3cm$, similar to an ErMnO$_3$ single crystal.[4] No secondary phases can be observed.

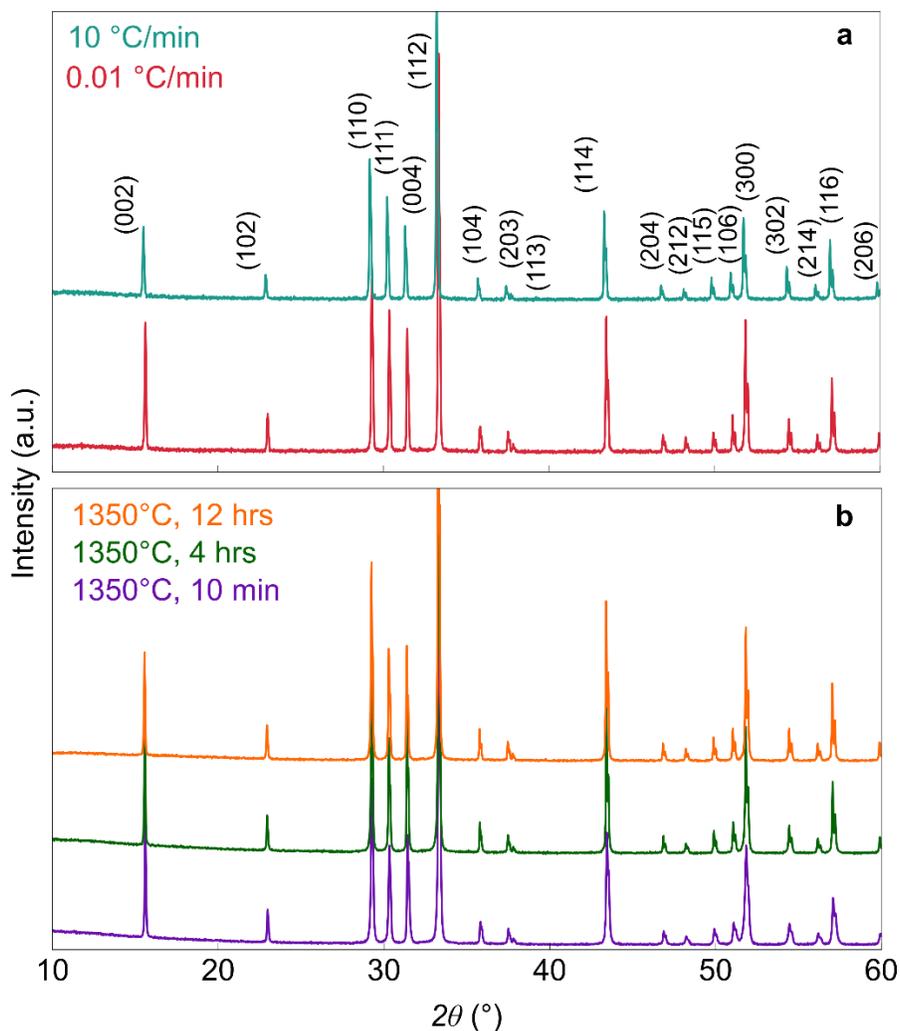

**Figure S1.** XRD patterns of polycrystalline pellets cooled under a) different rates (10 °C/min and 0.01 °C/min) and b) different heat-treatment conditions (1350°C, 10 min; 1350°C, 4 hrs; and 1450°C, 12 hrs). The data demonstrates that all samples have a hexagonal structure with space group symmetry $P6_3cm$.[4]

## S3. Analysis of micro- and domain structure

To visualize the micro- and domain structure of the samples, lapping was done with a 9-μm-grained Al$_2$O$_3$ water suspension (Logitech Ltd, Glasgow, Scotland) followed by polishing with silica slurry (SF1 Polishing Fluid, Logitech AS, Glasgow, Scotland).

Piezoresponse force microscopy (PFM) measurements were performed on an Ntegra Prisma system (NT-MDT, Moscow, Russia), with an electrically conductive platinum tip (Spark 150 Pt, Nu Nano Ltd, UK). The sample was excited using an alternating voltage (40.13 kHz, 10 V peak-to-peak). The laser deflection was read out by lock-in amplifiers (SR830, Stanford Research Systems, CA, USA). Topography data for the two different sample series is displayed in Fig. S2 (corresponds to PFM data displayed in Fig. 1) and Fig. S3 (corresponds to PFM data displayed in Fig. 3).

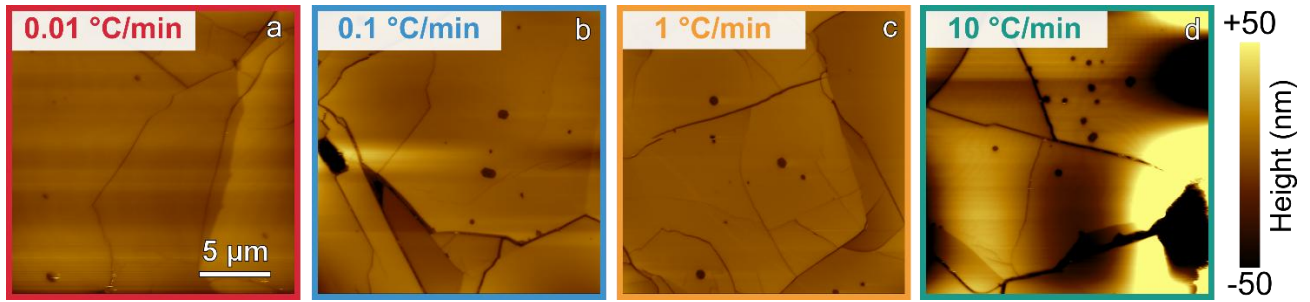

**Figure S2.** Topography data of the samples cooled at a) 0.01 °C/min, b) 0.1 °C/min, c) 1 °C/min, and d) 10 °C/min, covering the same area as shown for the PFM data in Fig. 1.

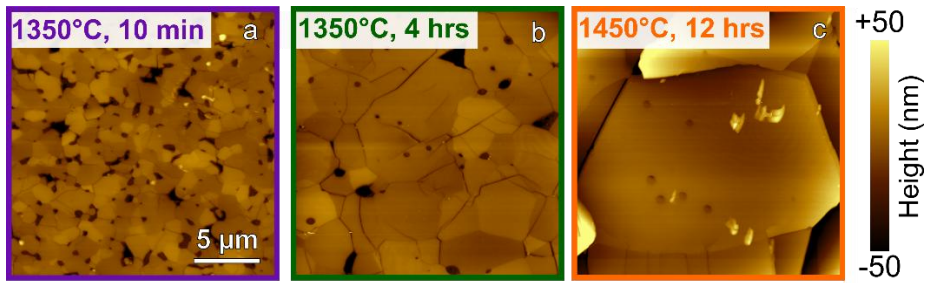

**Figure S3.** Topography data of the samples treated under different temperatures and dwell times, a) 1350°C, 10 min, b) 1350°C, 4 hrs, and c) 1450°C, 12 hrs, covering the same area as shown for the PFM data in Fig. 3.

The microstructure of the polycrystals, investigated by scanning electron microscopy (Tescan Mira 3, Czech Republic) operating in secondary electrons (SE) mode with an acceleration voltage of 5 kV is displayed in Fig. S4.

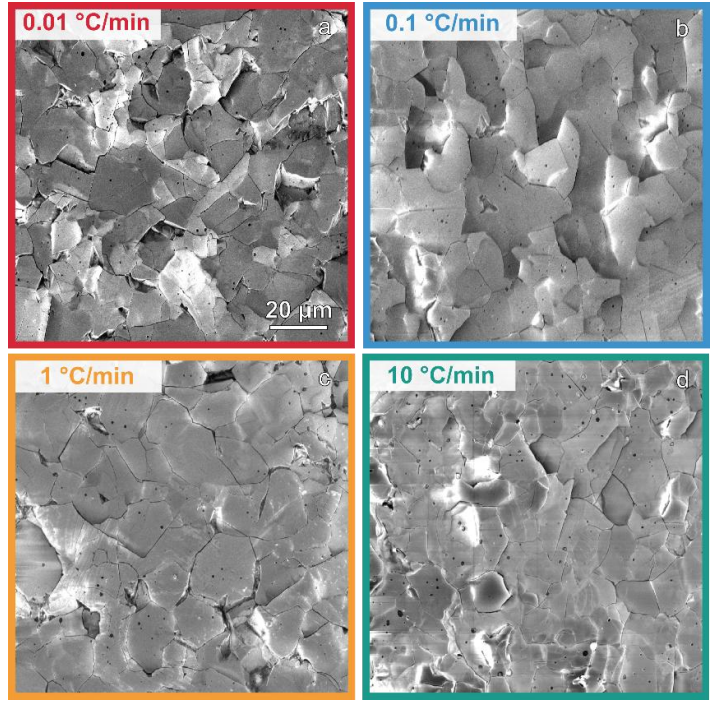

**Figure S4.** SEM micrographs of the samples cooled at a) 0.01 °C/min, b) 0.1 °C/min, c) 1 °C/min, and d) 10 °C/min.

### S4. Grain size and domain size determination

To obtain the domain size distribution, a stereographical method[5] is applied. The overall procedure, starting from the raw PFM data, is visualized for an exemplary grain in Fig. S5. First, for each grain, the shape of the grain boundary is manually determined based on the PFM image and the simultaneously recorded topography data (dashed line in Fig. S5a). Using the mean PFM signal within the entire grain as the threshold, a binarized image is calculated as shown in Fig. S5b.

Next, the domain walls are determined as the edges within the binarized image using a standard Canny edge algorithm[6] displayed in Fig. S5c. Subsequently, the distance of each pixel within the grain to the nearest pixel belonging to a domain wall (Fig. S5c) $d_{\text{p-dw}}$ is determined, leading after spatial averaging to an estimated mean domain size $d_{\text{estimate}}$ at each position as shown in Fig. S5d. The spatial averaging ensures that pixels close to a domain wall also recognize the distance to the next nearest domain wall and is calculated for each pixel $p$ as

$$d_{\text{estimate}}(p) = \max_{p' \in \text{grain}} \begin{cases} d_{\text{p-dw}}(p'), & \text{if } |p - p'| < 3 d_{\text{p-dw}}(p') \\ 0, & \text{else} \end{cases}. \quad (S1)$$

Based on the initial estimated domain size, the local mean value of the PFM signal within a circle with radius three times the estimated domain size is calculated for each pixel and applied as the threshold for local binarization in the second iteration, generating a new version of Fig. S5b. Following the previously described steps through the cycle leads to an updated local domain size (Fig. S5d) that is compared to the local domain size from the previous iteration. If the average estimated domain size changes by more than 1 ‰ within an iteration, another iteration is performed. After several iterative cycles, the estimated domain size saturates due to the achieved consistency between binarization and domain size estimation. This leads to the final representation of the domain walls shown in Fig S5e. The iterative sequence is required to account for local changes of the PFM background signal, e.g. due to topographic crosstalk, that cannot be captured by a global threshold value. Our algorithm builds up on existing approaches[7, 8] for variable-scale local thresholding developed in the field of image and document recognition and extends them towards the application of domain size determination in ferroelectrics.

After optimization, as shown in the inset of Fig. S5e intersections of straight lines with the domain walls (spacing of 470 nm between lines, repeated for 20 grids, each one rotated by 9° with respect to the previous one) and their respective distance are determined. This leads to a distribution of the lengths between the intersections, which we use for domain size quantification, as visualized in a histogram in Fig. S5f. Using Maximum-Likelihood optimization[9], the Gamma distribution *g(d)*, best fitting the measured distances *d* given by

$$g_{b,p}(d) = \frac{b^p}{\Gamma(p)} \cdot d^{(p-1)} \cdot \exp(-b \cdot d) \qquad (S2)$$

with the rate parameters *b* and the shape parameter *p* is determined. The probability distribution is chosen due to their application in Queueing theory on similar problems of sequentially occurring events with defined average probability of occurrence[10]. Finally,

$$\overline{d} = \int_0^\infty d' \cdot g_{b,p}(d') \mathrm{d}d' = p/b \qquad (S3)$$

and the respective uncertainty (due to the uncertainty of *b* and *p*) are used as the measure of the average domain size.

To calculate the mean grain size, the last part of the algorithm (after the last iterative cycle) is applied to the manually drawn outer shape of the grain instead of the final domain wall representation (Fig. S5e). Thereby the intersection points of the lines with the grain boundary can be directly calculated followed by the probability distribution-based calculation of the mean size.

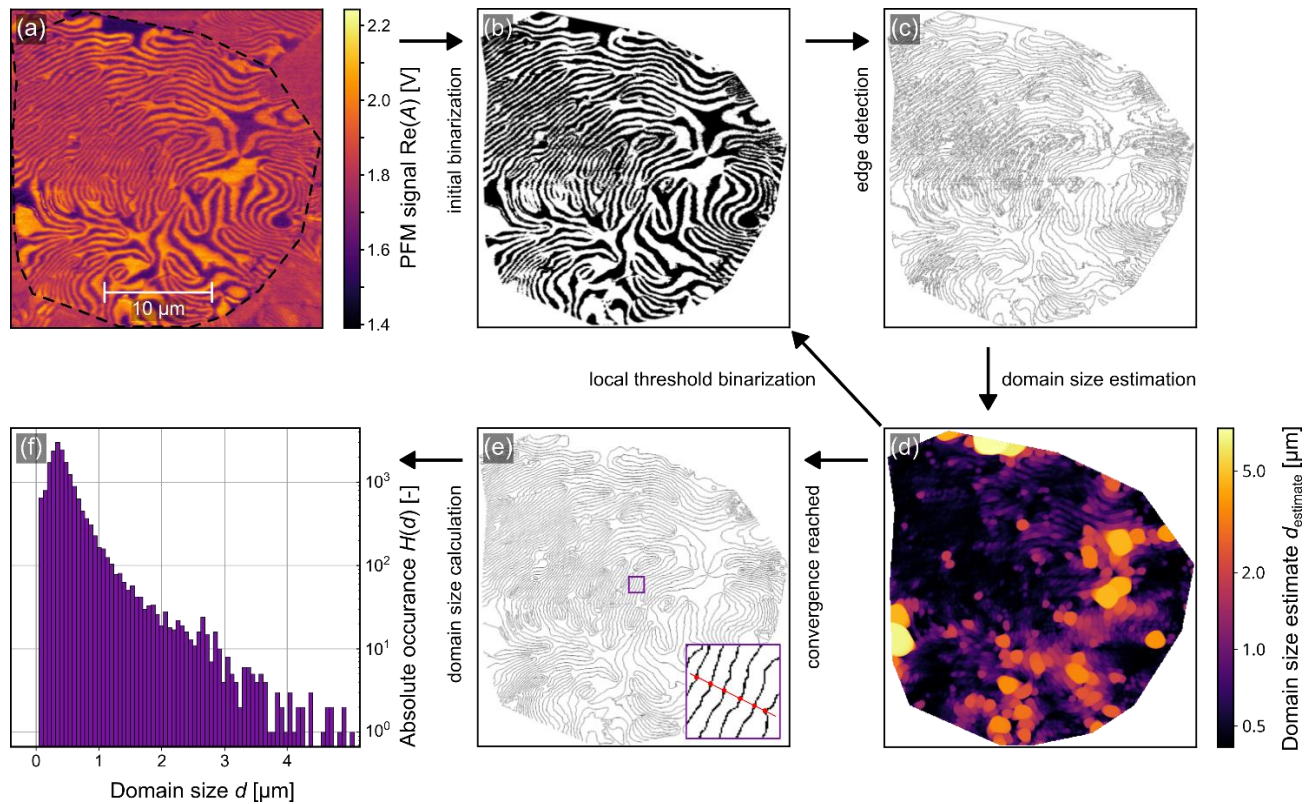

**Figure S5.** Steps of image processing algorithm for domain size determination. a) PFM image of a typical domain pattern with the dashed black line visualizing the grain boundary, b) binarized image, c) extracted domain walls, d) estimated local domain size, e) final extracted domain walls and f) final domain size distribution for the PFM image in a). The inset in e) shows the intersection of a straight line with the domain walls as applied in the final domain distance calculation step.

### S5. One-dimensional heat transfer model

Assuming a constant intrinsic thermal conductivity $\kappa_i$ for all samples, the thermal boundary resistances $R_{GB}$ and $R_{DW}$ are estimated for each temperature, following eq. 1 (Fig. S6a). The obtained mean values are $R_{GB} = 1.6 \cdot 10^{-9}$ m² K W⁻¹ and $R_{DW} = 2.8 \cdot 10^{-8}$ m² K W⁻¹. The corresponding intrinsic thermal conductivities $\kappa_i$ for each sample and temperature are shown in Fig. S6b. The variation in absolute values between the two samples series is attributed to differences in heat treatment conditions (see Supplementary Section S1 for more details). For each series, the standard deviation is in the range from 3 to 10%, confirming similar intrinsic thermal conductivity among the samples.

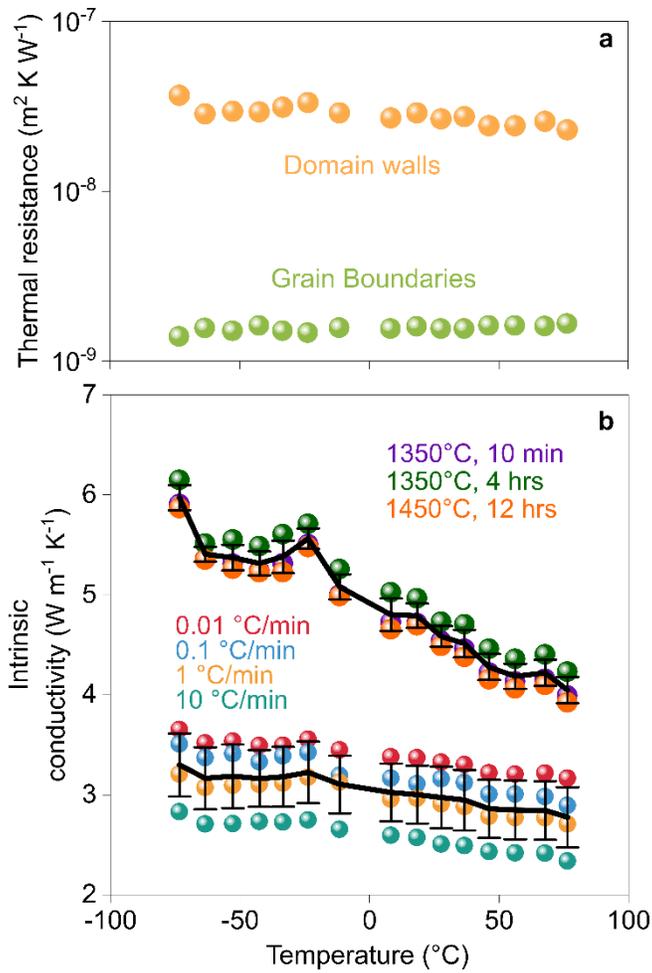

**Figure S6.** a) Estimated temperature-dependent thermal resistances, $R_{DW}$ and $R_{GB}$, for domain walls and grain boundaries, respectively. b) Temperature-dependent intrinsic thermal conductivity is calculated according to eq. (1) utilizing experimentally determined $\kappa$ values, grain and domain sizes from PFM data, and estimated thermal resistances. Black lines indicate the average value for each series, with standard-deviation displayed as error bars.